# The Limits of Motion Prediction Support for Ad hoc Wireless Network Performance

Stephen F. Bush
*GE Global Research*

Nathan Smith
*GE Global Research*

**Abstract**

*A fundamental understanding of gain provided by motion prediction in wireless ad hoc routing is currently lacking. This paper examines benefits in routing obtainable via prediction. A theoretical best-case non-predictive routing model is quantified in terms of both message overhead and update time for non-predictive routing. This best-case model of existing routing performance is compared with predictive routing. Several specific instances of predictive improvements in routing are examined. The primary contribution of this paper is quantification of predictive gain for wireless ad hoc routing.*

## 1. Introduction

Prediction of node location to support or supplant routing appears to be an attractive concept. Predictable episodic node location has been utilized for interplanetary communication [2]. It has been hypothesized that intuitively more chaotic node movement exhibited in terrestrial communication would benefit from predictable node location as well. However a fundamental study on the achievable gain in routing performance given perfect location prediction is required in order to better understand the benefits of predictability in network routing.

Existing routing algorithms compute the least-cost route at the current time. They do this with potentially stale information and without anticipating the possibility of better future routes. The risk of routing without prediction is choosing a candidate route that is likely to fail sooner than an alternate route and suffering the costs of a poor route along with route repair costs. Choosing a current higher-cost route, or waiting for a lower-cost one to form, may in fact be more beneficial than choosing the current least-cost route. The goal of this work is to recognize, and propose a model to quantify, the benefit versus cost of utilizing prediction.

The next section provides an analytical lower bound on current overhead and performance and considers the general relation between location prediction and motion entropy. Section 3 characterizes a tractable predictive routing model and assumptions used for exploring predictive gain. The model is used to examine four types of scenarios in which gains are hypothesized to exist in a predictive environment. Specific cases in which predictive routing gain is expected to be most apparent are examined. Section 4 summarizes the results of this work.

## 2. Predictive Routing Gain and Routing Overhead Reduction

One of the contributions of this paper is to consider the theoretical lower limit on route update overhead in a wireless environment. There is a plethora of ad hoc networking routing protocols with more routing protocols introduced periodically. A theoretical optimum routing performance analysis is necessary in order to benchmark how well routing protocols can ultimately be implemented. A lower limit on traditional routing overhead can be obtained by viewing motion as information whose entropy places a lower limit on compression and propagation of routing information. The next section develops the analysis in more detail.

### 2.1. Route Maintenance as Information Flow: Motion as the Information Generator

The minimum information required to encode routing information is quantified and then the optimal wireless network capacity is leveraged to propagate routing information. The result is a theoretical optimum routing protocol limit. Cluster areas are assumed in which nodes configure into clusters with a single cluster head. Cluster heads exchange neighbor information with one another in order to establish global routing. This provides a general representation of routing, for example, every node may be its own cluster or there may be multiple hierarchical levels of clustering. In the following analyses, all assumptions assume uniform random distributions unless specified otherwise; there are $N$ nodes within an area of size $A$ and there are $C$ cluster areas.

The information content in bits of to describe a local cluster, assuming the standard information measure based upon the probability of outcomes of an ensemble is shown in (1) [6].

$$I_c = \log_2 \binom{N}{\lceil N/C \rceil} \quad (1)$$

The more clusters that exist for a given number of nodes, the less information is required per cluster. However, more of this information has to be propagated to all other clusters. With a larger number of nodes, there may be more collective bandwidth that can be used for propagating the routing information. The maximum wireless capacity approximation [5] in a wireless broadcast media can be used to determine the collective capacity.

## 2.2. Propagating Route Update Information

Now that the minimum amount of information to be propagated has been determined, it must be propagated throughout the network as efficiently as possible. In this case, assume a perfect distribution mechanism in which all links are used as efficiently as possible to disseminate route update information. Assume a network of $n$ nodes is spread over an area $A$ and each possible connection has capacity $W$. Also, assume $\Delta$ is a guard distance to ensure channel transmissions do not overlap. The maximum wireless capacity in bit-meters per second is shown in (2).

$$C_{\max} = \sqrt{\frac{8}{\pi}} \frac{W}{\Delta} \sqrt{n} \quad (2)$$

Generalizing to a uniformly random distribution of $n$ sensors over a circular area $A$, the density is $\frac{n}{A}$, and the expected nearest-neighbor distance is $\frac{\sqrt{A}}{n}$. The total distance that data must travel is shown in (3).

$$d_{sum} = \sum_{k=1}^{n} \frac{\sqrt{A}}{n} \quad (3)$$

Putting the components together, the fundamental optimum bound for route update time is shown in (4). A single cluster results in a network in which all nodes are one hop away from a single cluster head; thus routing is through a single gateway and no inter-cluster propagation necessary. In the other extreme, the number of clusters increases and in the limit every node becomes its own cluster with simple information content but requiring propagation throughout the network. While the information per cluster is very low in this case, the information has to be propagated farther.

$$T_r = \frac{CI_c d_{sum}}{C_{\max}} \quad (4)$$

In a non-predictive routing protocol, route updates have a cost that can never reach this theoretical minimum due to inefficiencies in implementation. In order to facilitate the discussion at this point, consider that route update overhead, or cost as it is referred to throughout the reminder of this paper, is approximated as $C_r$. Note that route overhead cost may have units of either time $T_r$ from the equation in (4) or size $CI_c$ from the equation in (1). $C_r$ is comprised of elements such as the hello interval and its corresponding messages, route update intervals and corresponding messages, as well as other mechanisms that vary with the details of the specific non-predictive routing protocol implementation. For the predictive routing case, route update cost $C_p$ is a function of the entropy of node movement, or more specifically, the entropy of relative node movement $H_v$.

Legacy routing protocols exchange static tabular information. The goal of this effort is to lay the foundation for a mechanism where executable *models* of motion can be exchanged and the accuracy of the model determines the frequency at which routing updates are required. This paper seeks to explore under what circumstances $C_p(H_v) < C_r$. Predictive gain is then $G_p \approx C_r / C_p(H_v)$. Node motion causes a change in cluster membership and a corresponding generation of information $I_v$ that must be propagated to all nodes. Routing update information due to node motion is a function of relative node movement and the probability that a node changes connectivity from one cluster to another.

With regard to the impact node motion entropy, consider an absolute motion vector consisting of speed and direction for all nodes, $\vec{v}$. The curl (5) at various locations throughout the area of movement provides interesting information regarding irrotational locations [3].

$$\nabla \times \vec{v} \quad (5)$$

In general, predictive gain as a function of motion entropy is hypothesized to follow the curve shown in Fig. 1. Low motion entropy enables predictable location resulting in low routing overhead and high predictive gain. As motion entropy increases, location becomes less predictable and more location updates (update corrections) are required. Finally, as motion entropy becomes very high, precise location becomes unpredictable, however, overall node trajectory remains within a localized area and routing has the potential to become predictable.

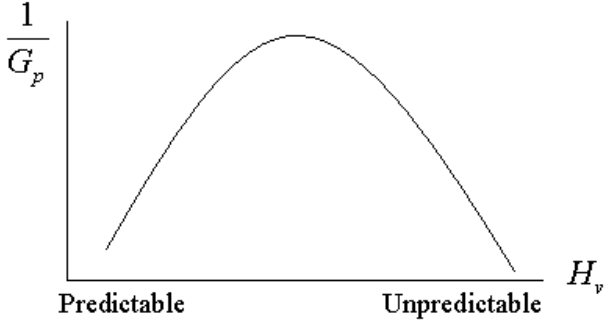

**Fig. 1. The expected trend in predictive gain is drawn as a function of motion entropy.**

A similar result regarding the relation of motion entropy to routing stability has been obtained in [1]. Specifically, nodes with high relative entropies yielded more stable links. A node velocity vector for node $n$ at time $t$ is shown in (6).

$$v(n,t) \qquad (6)$$

Relative node velocity vectors are shown in (7) between nodes $m$ and $n$.

$$v(m,n,t_i) = v(m,t) - v(n,t) \qquad (7)$$

The relative difference in velocity vectors between nodes $m$ and $n$ over $N$ samples is shown in (8).

$$a_{m,n} = \frac{1}{N}\sum_{i=1}^{N}|v(m,n,t_i)| \qquad (8)$$

The probability of a relative velocity of node $k$ is shown in (9). $F_m$ is a set of nodes of interest relative to node $k$. Note that (4) is an average speed over $N$ time samples; $\Delta_t$ is the total time over which the $N$ samples were taken.

$$P_{m,k}(t,\Delta_t) = \frac{a_{m,k}}{\sum_{i \in F_m} a_{m,i}} \qquad (9)$$

The entropy over set node set $F_m$ is shown in (10) where $C(F_m)$ is the number of nodes in the set relative to node $m$.

$$H_m(t,\Delta_t) = \frac{-\sum_{k \in F_m} P_k(t,\Delta_t) \log P_k(t,\Delta_t)}{\log C(F_m)} \qquad (10)$$

## 2.3. Motion Entropy and Topological Information Entropy

Node motion, ranging from random to highly coordinated, is generated as input to the motion entropy metric in (10). The goal is to generate a range of motion entropies from random to highly correlated and examine the impact on network topology and ultimately predictive capability. To create correlated node movement, a simple vector field is applied to obtain the coordinated node movement and random direction is used for random node movement. A variety of vector fields and super-position of vector fields have been explored; this one was chosen as a simple illustrative example. The ratio of nodes moving in random and vector field driven movement was varied to generate differing amounts of motion entropy. Fig. 2, Fig. 3, and Fig. 4 plot the relative motion entropy for a few of each node for ratios of 2:6, 3:5, 4:4, 5:3, and 6:2, of random to vector-driven nodes respectively. The resulting motion entropies plotted in Fig. 2 thru Fig. 4 correspond in every case to the relative ratios of random to coordinated nodes.

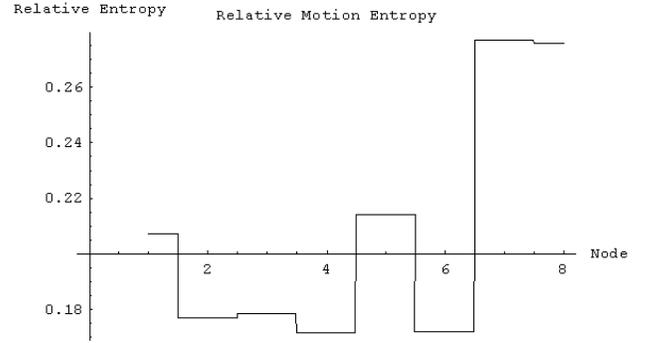

**Fig. 2. Motion entropy for two random and six vector guided nodes.**

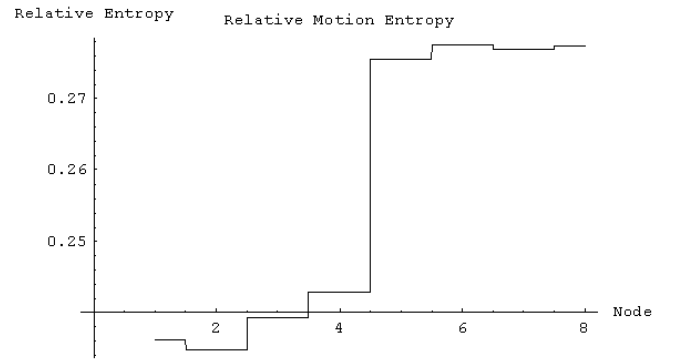

**Fig. 3. Motion entropy for four random and four vector guided nodes.**

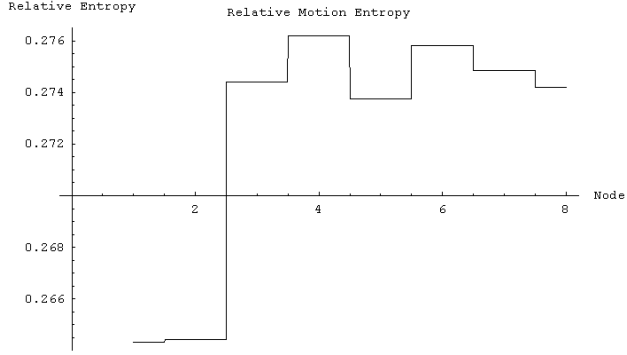

**Fig. 4. Motion entropy for six random and two vector guided nodes.**

Fig. 5 shows the corresponding inverse compression ratios of bit-strings that represent network topologies for random and coordinated node ratios of 1:7 (Motion Entropy 1 in Fig. 5), 2:6 (Motion Entropy 2), 3:5 (Motion Entropy 3), 4:4 (Motion Entropy 4), 5:3 (Motion Entropy 5), 6:2 (Motion Entropy 6), 7:1 (Motion Entropy 7) and includes results from each of the plots in Fig. 2 thru Fig. 4. The transmission area for each node was uniformly increased from 34% (Radius 1) to 75% (Radius 13) of the total field area in which the nodes were positioned. The network topology bit-string is formed such that each bit position represents a possible node connection for each of the $\binom{n}{2}$ possible connections. A bit set to one represents a connection and a bit set to zero indicates no connection. The inverse compression ratio forms the convex predictive gain curve hypothesized in Fig. 1 because both low and high motion entropies result in more compactly represented network topology bit-strings, while mid-level motion entropy results in more complex, less compact network topology bit-strings. In this case, the communication radius did not appear to have significant impact on the relation between motion entropy and the network topology bit-string representations.

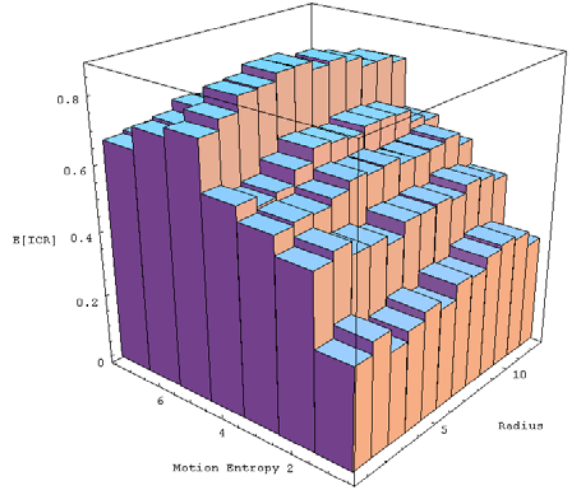

**Fig. 5. Expected inverse compression ratios of the network topology bit-string representation as a function of transmission radius and motion entropy.**

Regarding implementation, a distinction is made between optimistic and conservative predictive routing mechanisms. A conservative prediction mechanism transmits location updates to verify node location before link connectivity is attempted. An optimistic prediction mechanism assumes location prediction is perfect and attempts link connectivity without verification of node location. Link failure in the optimistic case forces a location verification and update. If location prediction is perfect, no overhead is required in the optimistic case and routing cost approaches zero ($C_p \to 0$). If prediction fails in all cases, then routing overhead reduces to non-predictive reactive routing costs ($C_p \approx C_r$).

## 3. Predictive Routing Model and Assumptions

Assuming uniformly random node movement that induces a random graph, the probability of link change is dependent upon expected node velocity and communication radius. The probability of a connection between two nodes is $(n-1)\pi r^2 / A$ where $A$ is the area in which nodes exist, $n$ is the number of nodes, and $r$ is the transmission radius. The rate connections change is the rate at which nodes leave or enter a new radius of communication area $\pi r^2$. If expected node velocity is $v$, then assuming uniformly random node movement, the node leaves the transmission area in $d/v$ time units, where $d$ is the expected distance to leave the covered area. Assume the mathematical model to be a random graph $G(n, p, w)$[1] with $n$ vertices, link connectivity probability $p$, and topology changes every $w$ time units.

In this model, the benefit of prediction can be determined by computing the likelihood that given an existing route of length $c_c$; a shorter route will form with the same source and destination node in less than $dt$ time units. The simplified routing model is used to quantify predictive routing gains. Predictive routing gain will be apparent when a shorter route is more likely to form than the current route, thus we quantify the probability of a route of a given length forming between two nodes in the graph $G(n, p, w)$.

With uniform random node movement, link formation with probability $p$ is shown in (11) where $r_{ij}$ is a route between nodes $i$ and $j$ of length $n_l$.

$$\Pr(r_{ij} \mid n, n_l) = \binom{n}{n_l} p^{n_l} (1-p)^{n-n_l} \quad (11)$$

### 3.1. A Faster Future Route

By waiting for a time $dt$, a route that supports faster transmission and completes an application flow in a shorter amount of time may come into alignment. The cost of using future route $c_s$ than one currently available $c_c$ may be less. The time waiting for the future route plus the transmission time using the future route may be less than the current fastest transmission route, in other words, a large inequality shown in (12) is desirable. With respect to time, there is both a continuous and discontinuous version of this approach. In this section the focus is on the continuous version; all links in a route are required to exist for data to flow. The predictive gain is shown in (13). As the application flow duration becomes smaller, approaching $dt$, the gain from the predicted route becomes less beneficial.

$$dt + c_s(t+dt) < c_c(t) \quad (12)$$

$$G_p = \frac{c_c(t)}{dt + c_s(t+dt)} \quad (13)$$

Consider a specific scenario in which 100 nodes with a communication radius of 2 meters in a $5000^2$ meter area. If there are currently more than 15 hops between a source and destination, then the likelihood is high that a shorter route will soon be available. In fact the probability that a route of less than 15 hops will form is 0.7. The geometric distribution whose mean is $p$ specifies the probability of successfully forming the shorter route given the number of graph changes, each taking time $w$. The expected value of (14) is $(1-p)/p$. Fig. 6 shows the probability of a route forming with less than 12 hops.

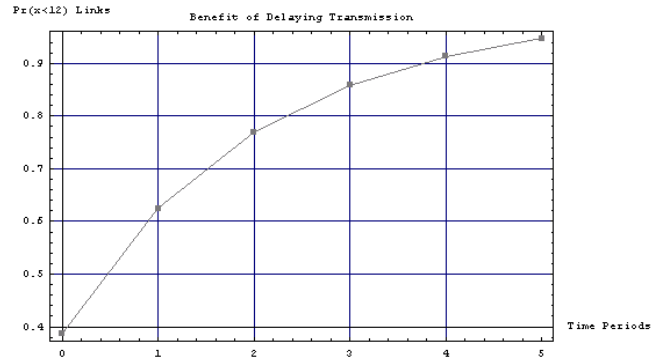

**Fig. 6. The probability of a route of fewer than 12 links forming is plotted as a function of the time taken for the route to form.**

If link changes occur every $w$ time units and the lifetime of an application flow is $f_t$ and each hop in the route reduces the effective bandwidth by $b_r$ bps, then the delay incurred per hop for a flow of duration $f_l$ is shown in (14).

$$d_f(n_l) = n_l f_l / b_r \quad (14)$$

By inserting (14) into (13), a differential gain is obtained shown in (15). It is the gain when the number of hops is currently $h$ and an attempt is made to wait for $h_{diff}$ fewer hops. $Et[h - h_{diff}]$ is the expected time for a shorter route to form. The gain increases as the likelihood of a shorter route increases, reaching a maximum, and then decreasing as the likelihood of a shorter route decreases. A larger $h_{diff}$ indicates a shorter route and a shorter route would have a corresponding improvement in gain. However, the likelihood of short routes becomes lower at high values of $h_{diff}$. The maximum absolute gain is plotted in Fig. 7 as a function of the per-hop delay $b_r$ exponent. Large effective delays per hop increase the value of shorter routes and improve gain. The results are plotted for a 100-node network in a $5000^2$ meter area with a radius of communication of 2 meters, a transmission rate to 1 Mbs, a transmission of $10^8$ bits ($f_l = 10^8$), a network topology change every 10

---
[1] The use of a random graph does not perfectly represent node movement, which is likely to be conditional upon last location. However, the random graph provides a tractable framework for illustrating the predictive concepts.

seconds ($w=10$), and a starting hop count of $h=50$ hops.

$$G_{dp}(h|h_{diff}) = \frac{c_c(h)}{d_f(h-h_{diff}) + Et[h-h_{diff}]} \quad (15)$$

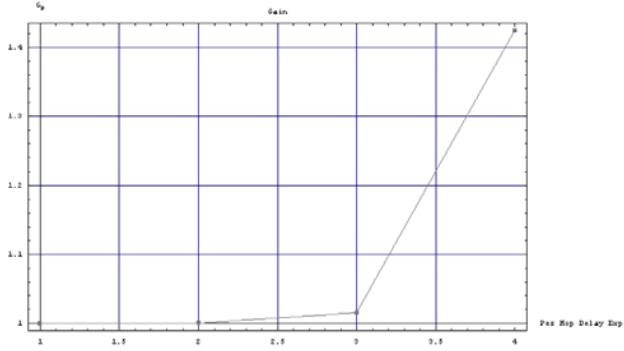

**Fig. 7. Absolute gain is plotted as a function of the per-hop delay exponent.**

### 3.2. Discovering Flash Routes

The term *flash route*[2] is introduced in this paper as a route from a source to a destination node, whose duration is too short, using non-predictive routing protocols, to be useful or perhaps even detectable. However, using predictive routing techniques, flash routes may be leveraged. The concept is that location prediction indicates that a route is likely to exist, particularly a route with a short lifetime. In fact, the lifetime may be so short that required route updates may not propagate to the source or destination of the route before the route disappears. Transmission optimistically takes place over the predicted route even though traditional routing tables would not have time to update using traditional routing techniques.

Using the simplified $G(n,p,w)$ model proposed in this paper, a flash route exists when route update time $T_r$ is greater than the lifetime of a route. Since route updates must follow existing links at the current time, route updates will not be able to indicate routes that disappear before updates are received. From (4), the theoretical fastest route update time can be estimated. The probability that a route exists whose length from source to destination is less than a given number of hops and whose duration is *less* than the time required to be detectable via a best-case route update (4) must be quantified.

Given that $p_{n_l}$ is the expected probability for a route of length $n_l$ to form (11), then (20) is the expected time for a continuous route of length $n_l$ to form. A long route that forms quickly has a higher likelihood that it will form and disappear before the best-case route update can complete to notify the network of its existence. A route will disappear with likelihood $n_l(1-p)$ and the time for the theoretically most efficient route update for a route of a given number of hops can be derived from (3) and (4) and is represented below as $T_{r_{n_l}}$. Inequality (21) is the expected likelihood of a route of length $n_l$ disappearing before an optimal route update can propagate the length of the route.

$$\frac{1-p_{n_l}}{p_{n_l}} w \quad (20)$$

$$p_{n_l} n_l (1-p) w < T_{r_{n_l}} \quad (21)$$

In Fig. 8 the probability of the existence of a flash route as a function of route length and graph change interval for the previous case of a 100-node network in a $5000^2$ meter area with a radius of transmission of 2 meters is shown. As node motion increases the likelihood of a flash route increases with the length of the potential flash route. A flash route must be long enough such that it would take too long for a route update to notify the network of its existence and node motion must be large enough that the route is likely to exist for only a short duration. Motion prediction allows the capability of leveraging this large capacity of previously unavailable routes.

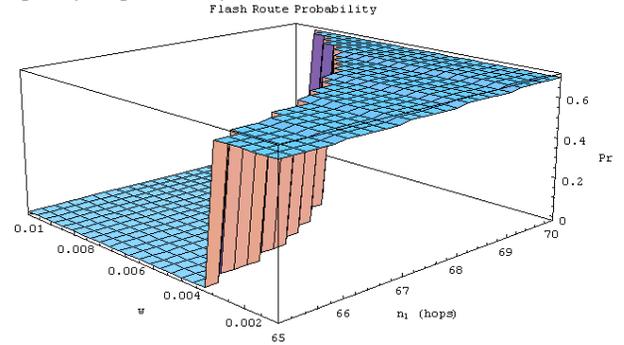

**Fig. 8. Probability of a flash route is plotted as a function of flash route length and graph topology change interval ($w$).**

## 4. Conclusion

This paper has examined and quantified selected benefits in routing that are obtainable with prediction. Several specific cases of predictive improvements in routing are examined; namely, delaying to improve transmission time, robust routing around mobile noise sources, temporally discontinuous routing, and the harvesting of flash routes, which are a new form of predictive routing gain defined in this paper. These results indication that there is significant gain in network routing to be derived via improvements in location prediction.

---
[2] The primary author, while exploring benefits of predictive capabilities, coined the term "flash route" with the intuitive meaning of very short duration routes.